\documentclass{revtex4} \usepackage{latexsym}
\usepackage{amsfonts} \usepackage{amsmath,revsymb}                        

\begin{document}

\newcommand{\HS}{\mathcal{H}}
\newcommand{\kt}[1]{|#1\rangle}
\newcommand{\vp}{\mathbf{p}}
\newcommand{\vP}{\mathbf{P}}
\newcommand{\vq}{\mathbf{q}}

\title{Tensor Product Structures, Entanglement,\\ and Particle Scattering}

\author{N.L.~Harshman\\{\footnotesize\it Department of Computer Science, Audio Technology, and Physics, American University,\\ 4400 Massachusetts Ave.\ NW, Washington, DC, 20016, USA, harshman@american.edu}\\[2ex]
 S.~Wickramasekara\\{\footnotesize\it Department of Physics, Grinnell College, Grinnell, IA 50112, USA, wickrama@grinnell.edu}}

\begin{abstract}
  Particle systems admit a variety of tensor product structures (TPSs) depending on the complete system of commuting observables chosen for the analysis.  Different notions of entanglement are associated with these different TPSs.  Global symmetry transformations and dynamical transformations factor into products of local unitary operators with respect to certain TPSs and not with respect to others.  Symmetry-invariant and dynamical-invariant TPSs and corresponding measures of entanglement are defined for particle scattering systems.
\end{abstract}
\maketitle

\section{Introduction}

In this paper, we apply the notion of tensor product structures (TPSs) to dynamical entanglement in particle scattering.  A major motivation for this work is the significance of scattering as a main experimental paradigm in physics. Many quantum measurements, especially in high energy physics,  can be couched in the language of scattering. Scattering is also the dominant dynamic mechanism of some many-body systems, an important fact in view of the recently established connection between entanglement and criticality in many-body systems (see, for example, in this volume \cite{anders}). Also, the methods of analysis of  particle scattering often carry over into new physical regimes by way of the scattering of quasiparticles or other types of excitations.

Within the context of quantum information theory, there are many open questions about entanglement in systems with continuous degrees of freedom (DOF) or both continuous and discrete DOF.  The main focus of this  work is the qualitative notion of separability with respect to a particular TPS as relevant for systems with continuous DOF. One of the main results that we report is that by considering a physical system with known symmetry properties, like non-relativistic or relativistic particles, some results about separability,  the dynamical generation of entanglement, and the classification of entangling operators can be obtained without the computation of specific entanglement monotones.

Despite the central significance of particle scattering processes to our understanding of physics, there are surprisingly few studies on the entanglement properties of scattering particles.  Most of these works have treated only special cases or specific physical systems and advances toward a complete, general theory have been meager.  We see the absence of a consensus about the relevance of various entangling DOF as one major hindrance to overcome before achieving generality.  See, for example, the literature on spin-entanglement of relativistic particles~\cite{relent, goodrel} where different types of entanglement (between two particles, between two particles' spins, between two particle's momenta, and between a single particle's spin and momentum) have been discussed and occasionally confused.  For other results for entanglement in scattering, see, for example, the references \cite{momenta} which consider interparticle momentum entanglement and \cite{spin} which consider interparticle spin or angular momenta entanglement. We view symmetry arguments as a means of categorizing various notions of entanglement and thereby paving the way toward a general theory of entanglement, especially dynamical entanglement,  in particle scattering.  

In this paper we explore TPSs appropriate for particle scattering, and in particular highlight the usefulness of symmetry-invariant and dynamical-invariant TPSs.  Notions of separability and entanglement are TPS dependent, and consequently have different transformation properties under symmetry operations.  We consider it a primary criterion that any useful entanglement measure must have well defined transformation properties under an appropriate spacetime symmetry group (i.e., Galilei or Poincar\'e) and possibly also a gauge group. We further remark that some authors prefer to use the term entanglement only for interparticle entanglement and to call any other kind of entanglement ``generalized entanglement''. This accords interparticle entanglement special prominence. Insofar as scattering dynamics and transformation properties under symmetry operators are concerned, there is nothing fundamentally different or significant about interparticle entanglement albeit it is perhaps the one that appertains to the initial conditions of the typical scattering in-states. Thus we will simply use the term entanglement to refer to that between any suitable set of DOF. 

Among the outcomes of the concepts outlined above, the invariance of entanglement between the internal DOF and external DOF (defined precisely below) of a multi-particle system in scattering dynamics is particularly interesting.  This result is also found in \cite{harshwick1}, but here we provide some details, an alternate proof, and other examples.

\section{Tensor Product Structures: An Example with Qubits}

Several works provide an introduction to the idea of TPSs \cite{zanardi,zanlidar,barnum}.  In this section we expand an example from \cite{zanlidar} to show the relationships between TPSs, observables, entanglement, and symmetry representations in the familiar context of qubits.

Consider two qubits, $A$ and $B$, whose states are represented by elements of two separate Hilbert spaces $\HS_A$ and $\HS_B$, respectively.  These Hilbert spaces are isomorphic and can be realized as $ \mathbb{C}^2$:
\begin{equation}
\HS_A \sim \HS_B \sim \mathbb{C}^2.
\end{equation}
The Hilbert space $\HS$ for the combined system is the tensor product of $\HS_A$ and $\HS_B$, which is isomorphic to $\mathbb{C}^4$:
\begin{equation}\label{TPS:ipqb}
\HS = \HS_A \otimes \HS_B \sim \mathbb{C}^4.
\end{equation}

With respect to the TPS (\ref{TPS:ipqb}), which we will refer to as AB-TPS, the  individual qubit algebras  of observables have an embedding as $\mathcal{A}_A\otimes{\mathbb{I}}_B$ and ${\mathbb{I}}_A\otimes\mathcal{A}_B$ in the algebra of operators in $\HS$. 
\begin{eqnarray}
\mathcal{A}_A\otimes{\mathbb{I}}_B &=& \mathrm{span}_\mathbb{R}\{\mathbb{I}_A\otimes\mathbb{I}_B, \hat{\sigma}^x_A\otimes\mathbb{I}_B,\hat{\sigma}^y_A\otimes\mathbb{I}_B,\hat{\sigma}^z_A\otimes\mathbb{I}_B\}\\
{\mathbb{I}}_A\otimes \mathcal{A}_B &=& \mathrm{span}_\mathbb{R}\{\mathbb{I}_A \otimes \mathbb{I}_B, \mathbb{I}_A \otimes \hat{\sigma}^x_B, \mathbb{I}_A \otimes \hat{\sigma}^y_B, \mathbb{I}_A \otimes \hat{\sigma}^z_B\},
\end{eqnarray}
where $\mathbb{I}_A$ and $\hat{\sigma}^i_A$ are the identity operator and Pauli matrices for the qubit $A$ and  $\mathbb{I}_B$ and $\hat{\sigma}^i_B$ are the identity operator and Pauli matrices for the qubit $B$.  The algebras $\mathcal{A}_A$ and $\mathcal{A}_B$ are the enveloping (associative) algebras of the $\mathrm{u}(2)$ Lie algebra for qubits $A$ and $B$, respectively.  When the meaning is clear, we will often denote the embedded algebras $\mathcal{A}_A\otimes{\mathbb{I}}_B$ and 
${\mathbb{I}}_A\otimes\mathcal{A}_B$ by the simpler notation $\mathcal{A}_A$ and $\mathcal{A}_B$.  
The algebras ${\mathcal{A}}_A$ and ${\mathcal{A}}_B$ commute with each other, the mathematical expression that the measurements of the properties of $A$ and $B$ are independent.  In fact, it is the commutativity (and completeness) of these subalgebras that gives credence to the AB-TPS for the combined $AB$ system. 

A complete system of commuting observables (CSCO) consists of any one non-trivial element from each subalgebra $\mathcal{A}_A$ and $\mathcal{A}_B$.
For instance, we may choose $\{\hat{\sigma}^z_A\otimes{\mathbb{I}}_B,{\mathbb{I}}_A\otimes\hat{\sigma}_B^z\}$.  The four vectors that diagonalize this CSCO furnish a basis for $\HS$ that is adapted to the TPS (\ref{TPS:ipqb}). Using the conventional two two-valued indices $j, k \in \{ 0, 1\}$, the eigenvectors for the CSCO 
$\{\hat{\sigma}^z_A\otimes{\mathbb{I}}_B,{\mathbb{I}}_A\otimes\hat{\sigma}_B^z\}$ can be labeled as $|jk\rangle$ such that, for instance, 
$ (\hat{\sigma}^z_A\otimes\mathbb{I}_B)\kt{01} = -\kt{01}$ and $(\mathbb{I}_A \otimes \hat{\sigma}^z_B)\kt{01} = \kt{01}$.
This basis is a realization of the AB-TPS. 

Any state $|\phi\rangle\in\HS$ can be written as a linear combination of the $|jk\rangle$:
\begin{equation}
\kt{\phi} = \kt{\phi_A}\otimes\kt{\phi_B} = \sum_{j,k}a_j b_k \kt{jk}.
\end{equation}
Such a general normalized state, separable or entangled, can also be expressed by the singular value decomposition as
\begin{equation}
\kt{\phi} = \sum_{i=1}^2 \lambda_i \kt{\psi_i},
\end{equation}
where the $\kt{\psi_i}$, $i\in\{1,2\}$, are orthonormal and separable with respect to the AB-TPS (\ref{TPS:ipqb}) and $\lambda_1^2 + \lambda^2_2 = 1$.  
The entropy of entanglement for the AB-TPS (\ref{TPS:ipqb}) is
\begin{equation}
E_AB(\phi) = - \sum_{i=1}^2 \lambda_i^2 \log_2\lambda_i^2\label{entropy}
\end{equation}

Other TPS are possible for the Hilbert space $\HS$ of the combined AB-system. To that end, following Zanardi and Lidar~\cite{zanlidar}, let us define four state vectors $\kt{\chi\xi}$, $\chi,\xi=\{+,-\}$ which have maximum entanglement with respect to ({\ref{entropy}}):
\begin{eqnarray}
\kt{+,\pm} &=& \frac{1}{\sqrt{2}} \left( \kt{00} \pm \kt{11} \right)\\
\kt{-,\pm} &=& \frac{1}{\sqrt{2}} \left( \kt{01} \pm \kt{10} \right).
\end{eqnarray}
These four state vectors are eigenvectors of the 
CSCO $\{\hat{\sigma}^z_A\otimes\hat{\sigma}^z_B, \hat{\sigma}^x_A\otimes\hat{\sigma}^x_B\}$ and thus 
constitute a basis for $\HS$:
\begin{eqnarray}
(\hat{\sigma}^z_A\otimes\hat{\sigma}^z_B)\kt{\chi\xi} &=& \chi\kt{\chi\xi}\\
(\hat{\sigma}^x_A\otimes\hat{\sigma}^x_B)\kt{\chi\xi} &=& \xi\kt{\chi\xi}
\end{eqnarray}
This observation leads to a new way of constructing a TPS for $\HS$. We define the algebras of observables $\mathcal{A}_P$ and $\mathcal{A}_Q$:
\begin{eqnarray}
\mathcal{A}_P&=& \mathrm{span}_\mathbb{R}\{\mathbb{I}_A\otimes\mathbb{I}_B, \hat{\sigma}^x_A\otimes\mathbb{I}_B,\hat{\sigma}^y_A\otimes\hat{\sigma}^z_B,\hat{\sigma}^z_A\otimes\hat{\sigma}^z_B\}\label{P}\\
\mathcal{A}_Q &=& \mathrm{span}_\mathbb{R}\{\mathbb{I}_A \otimes \mathbb{I}_B, \mathbb{I}_A \otimes \hat{\sigma}^z_B, \hat{\sigma}^x_A \otimes \hat{\sigma}^y_B, \hat{\sigma}^x_A \otimes \hat{\sigma}^x_B\}\label{Q}
\end{eqnarray}
These algebras are isomorphic to each other, and to the original algebras ${\mathcal{A}}_A$ and ${\mathcal{A}}_B$.  They commute and represent another way to partition the observables of the physical system described by $\HS$.  In particular, the algebras $\mathcal{A}_P$ and $\mathcal{A}_Q$ have realization as algeras of bounded operators in two Hilbert spaces $\HS_P$ and $\HS_Q$, each two dimensional, such that the total Hilbert space $\HS$ for the two qubit system is the direct product $\HS_P\otimes\HS_Q$. That is, the algebras (\ref{P}) and (\ref{Q}) induce a new tensor product structure, which we call PQ-TPS:
\begin{equation}
\HS_P \sim \HS_Q \sim \mathbb{C}^2
\end{equation}
and
\begin{equation}\label{TPS:pqqb}
\HS = \HS_P \otimes \HS_Q \sim \mathbb{C}^4
\end{equation}
Separability and entanglement defined with respect to the PQ-TPS are different from those defined with respect to the AB-TPS.  In particular,  the singular value decomposition for the PQ-TPS will have Schmidt coefficients $\tilde{\lambda}_i$ that are different from the $\lambda_i$ of the AB-TPS, 
and the same state $|\phi\rangle$ will generally have a different amount of  entropy of entanglement depending on the underlying TPS used for the singular value decomposition.  In fact, the $\kt{jk}$ basis vectors of the AB-TPS are maximally entangled in the PQ-TPS, thereby completing the reciprocal construction.  
The moral of this story is that the observables determine the TPS, which in turn determines the notions of separability and entanglement.

\section{Free Particle TPSs}

The two tensor product structures AB-TPS and PQ-TPS that we constructed for the two qubit system is were equivalent in the sense that each TPS factors the total Hilbert space $\HS\sim{\mathbb{C}}^4$ into $\mathbb{C}^2\otimes\mathbb{C}^2$. The only non-trivial factorization of ${\mathbb{C}}^4$ is 
$\mathbb{C}^2\otimes\mathbb{C}^2$ and thus all TPSs for the two qubit system are equivalent as they induce the same factorization of the total Hilbert space $\HS$.  
There exist different non-trivial factorizations of higher dimensional Hilbert spaces, and therewith non-equivalent TPS. In this section, we consider the momentum and spin DOF of  non-relativistic particles. The Hilbert spaces are now infinite dimensional and have many non-equivalent partitions.  Another way to say this is that there exist different, inequivalent  CSCO in the algebra of observables for a non-relativistic particle. 

The algebra of observables for a single non-relativistic particle is the enveloping algebra of the Galilean Lie algebra.   In quantum physics, what is relevant is Lie algebra of (the covering group of) $\mathcal{G}$, the Galilean group in $3+1$ dimensions, extended by the central charge of mass.  We denote the elements of $\mathcal{G}$ by $g=(b,{\bf a},{\bf v},R)$, where $b\in\mathbb{R}$ is a time translation, ${\bf a}\in\mathbb{R}^3$ a space translation, ${\bf v}\in\mathbb{R}^3$ a velocity boost, and $R\in\mathrm{SO}(3)$ (or $u\in\mathrm{SU}(2)\rightarrow R(u)\in\mathrm{SO}(3)$, the standard 2-to-1 homomorphism) is a rotation.  The associated basis for the operator Galilean algebra in the Hilbert space of states is $\{\hat{H},\hat{\mathbf P},\hat{\mathbf Q}, \hat{\mathbf J}\}$.  The mass $\hat{M}$ can be added as a central element of this operator Lie algebra. The mass-extended enveloping algebra includes the position operator  $\hat{\mathbf X}=\hat{\mathbf Q}\hat{M}^{-1}$, the orbital angular momentum vector operator $\hat{\mathbf L}=\hat{\mathbf X}\times\hat{\mathbf P}$, and the intrinsic spin vector operator $\hat{\mathbf S}=\hat{\mathbf J}-\hat{\mathbf L}$.  The operators corresponding to internal energy $\hat{W}=\hat{H} - (2\hat{M})^{-1}\hat{\mathbf P}^2$, intrinsic spin squared $\hat{\mathbf S}^2$ and (trivially) mass $\hat{M}$ are the Casimir operators, i.e., they commute with the entire enveloping algebra. In unitary irreducible representations (UIR) of the Galilean group, the Casimir operators are all proportional to the identity operator,  $\hat{M}=mI,\ \hat{W}=WI,\ \hat{\mathbf S}^2=s(s+1)I$, and parameters $m$, $W$ and $s$ of these eigenvalues characterize the representation Hilbert spaces $\HS(m,W,s)$ ~\cite{levy}.  In the single particle case, it is possible to set  $W=0$ without loss of generality.  While they characterize the representation space for a particular non-relativistic quantum particle, the Casimir operators  are not relevant to discussing inequivalent partitions of the single particle Hilbert space.

The operators that label the degeneracy within the single particle Hilbert space can be chosen in many ways.  For example, a standard choice is the momentum operators $\{\hat{P}_x, \hat{P}_y, \hat{P}_z\}$ and one spin component operator, say $\hat{S}_z$.  The tensor product implied by this choice of CSCO is
\begin{equation}\label{TPS:3msop}
\HS_{p_x} \otimes \HS_{p_y} \otimes \HS_{p_z} \otimes\HS_{s} \sim \mathrm{L}^2(\mathbb{R}) \otimes \mathrm{L}^2(\mathbb{R}) \otimes \mathrm{L}^2(\mathbb{R}) \otimes \mathbb{C}^{s(s+1)}.
\end{equation}
This tensor product of four Hilbert spaces can be partitioned into seven 2-way splits (of two inequivalent kinds), six 3-way splits and one 4-way split.  One partition of physical interest for general scattering problems is the bipartite TPS
\begin{eqnarray}\label{TPS:msop}
\left(\HS_{p_x} \otimes \HS_{p_y} \otimes \HS_{p_z}\right) \otimes\HS_{s} &=& \HS_\mathbf{p} \otimes \HS_{s}\\
&\sim& \mathrm{L}^2(\mathbb{R}^3) \otimes \mathbb{C}^{s(s+1)}
\end{eqnarray}
that separates the momentum Hilbert space from the spin Hilbert space.

Other one-particle CSCOs lead to other TPSs.  The set of TPSs can be induced from the position and spin CSCO $\{\hat{R}_x, \hat{R}_y, \hat{R}_z, \hat{S}_z\}$ is equivalent to (\ref{TPS:3msop}) and its partitions because the Fourier transform of the momentum operators and functions spaces is a local unitary operator with respect to (\ref{TPS:3msop}).  One could also imagine other mixed possibilities of some position and some momentum operators, but these TPS structures would also be unitarily equivalent.

Tensor product structures not equivalent to the above exist. For example, consider the CSCO $\{\hat{H}, \hat{L}^2, \hat{L}_z, \hat{S}_z\}$, where $\hat{H}$ is the free single particle Hamiltonian and $\hat{L}$ is the orbital angular momentum. This CSCO induces the TPS
\begin{equation}\label{TPS:elsop}
\HS_E\otimes \HS_{L} \otimes \HS_s \sim \mathrm{L}^2(\mathbb{R}^+) \otimes \mathrm{L}^2(\mathbb{S}^2) \otimes \mathbb{C}^{s(s+1)},
\end{equation}
where $\mathrm{L}^2(\mathbb{S}^2)$ are Lebesgue square-integrable functions on the 2-sphere and
\begin{equation}
\mathrm{L}^2(\mathbb{S}^2)\sim \bigoplus_l \mathbb{C}^{l(l+1)}.
\end{equation}
This tripartite TPS is not equivalent to (\ref{TPS:msop}), although there is a natural isomorphism from the space $\HS_\mathbf{p}$ to the space $\HS_E\otimes \HS_{L}$ induced by the coordinate change from rectangular to spherical coordinates.

As another possibility in the same vein, the product $\HS_{L} \otimes \HS_s$ can be reduced into a direct sum of HSs using Clebsch-Gordan methods for the rotation group, and this is equivalent to using the TPS associated with the total angular momentum CSCO $\{\hat{H}, \hat{L}^2, \hat{J}^2, \hat{J}_z\}$
\begin{eqnarray}\label{TPS:ejop}
\HS_E\otimes \HS_J &=& \HS_E \otimes \left(\bigoplus_{j} \HS_j \otimes \HS_{j,l,s}\right)\\
&\sim& \mathrm{L}^2(\mathbb{R}^+) \otimes \left(\bigoplus_{j} \mathbb{C}^{j(j+1)} \otimes \mathbb{C}^{d(j,s)}\right),
\end{eqnarray}
where $d(j,s)$ is the number of times $\HS_j$ appears in the decomposition of $\HS_{L} \otimes \HS_s$.  This degeneracy is labeled by orbital angular momentum $l$.

Other single-particle TPSs are possible, but, to summarize, three CSCOs common in non-relativistic particle physics define the momentum-spin (or, equivalently, position-spin) bipartite TPS (\ref{TPS:msop}), the energy-orbital-spin angular momentum tripartite TPS (\ref{TPS:elsop}), and the energy-total angular momentum bipartite TPS (\ref{TPS:ejop}).  With each of these TPSs, there is a different notion of separability, and therefore a different kind of entanglement.  This is entanglement between DOF, not between ``particles'', so this kind of entanglement does not bring up non-locality issues.  Nonetheless, this kinds of intraparticle entanglement can evolve, participate in quantum information processes, and in principle be converted to the standard qubit entanglement via controlled interactions and measurement.

Next, let us consider TPSs for two non-relativistic particles, in particular two particles that interact via scattering, taken to be elastic for the sake of simplicity.  If A and B are two such particles, the Hilbert space for the two particle states is the direct product $\HS=\HS_A \otimes \HS_B$, where $\HS_A$ and $\HS_B$ are Hilbert spaces that furnish UIRs of the Galilean group characteristic of the particles A and B.  The A and B begin and end the experiment outside of the interaction region, and so the TPS pertaining to the factorization $\HS=\HS_A \otimes \HS_B$ is well-suited for describing the  in-state and out-state of the system. This TPS, which we call the interparticle TPS, is the TPS that leads to the kind of interparticle entanglement that is typically studied.  The boundary conditions of scattering imply that there is no interpartical entanglement before scattering, and but scattering dynamics typically change this, leading to an entangled out-state.

Clearly, many other TPSs exist for the two particle system.  Some can be constructed by inheriting the one-particle TPSs and then repartitioning.  Some examples include 
\begin{equation}
\HS_{\mathbf{p}_{tot}}\otimes\HS_{s_{tot}} = \left(\HS_{\mathbf{p}_A} \otimes \HS_{\mathbf{p}_B}\right) \otimes \left(\HS_{s_A}\otimes \HS_{s_A}\right),
\end{equation}
which could be used to investigate the entanglement between the total momentum and total spin DOF,
and
\begin{equation}
\HS_{\mathbf{E}_{tot}}\otimes\HS_{J_{tot}} = \left(\HS_{\mathbf{E}_A} \otimes \HS_{\mathbf{E}_B}\right) \otimes \left(\HS_{J_A}\otimes \HS_{J_A}\right),
\end{equation}
which partitions the total energy and total angular momentum.

Other  TPSs exist for the two particle case that are not just repartitioning of the one particle TPSs.  For example, consider two spinless particles. The change of variables to total and relative momentum
\begin{equation}\label{com}
\vP =\vp_A + \vp_B, \quad
\vq = \frac{1}{m_A + m_B} (m_B\vp_A - m_A\vp_B)
\end{equation}
gives rise to the unitary transformation of the Hilbert space 
\begin{eqnarray}\label{tps-tp4}
\HS_{\vp_A} \otimes \HS_{\vp_B} &\rightarrow& \HS_{\vp} \otimes \HS_{\vq}\nonumber\\
\mathrm{L}^2(\mathbb{R}^3)\otimes\mathrm{L}^2(\mathbb{R}^3)\otimes \ &\rightarrow& \mathrm{L}^2(\mathbb{R}^3)\otimes\mathrm{L}^2(\mathbb{R}^3).
\end{eqnarray}

The transformation of variables (\ref{com}) and of TPS (\ref{tps-tp4}) is the first step in finding the Clebsch-Gordan series for the reduction of the direct product of unitary irreducible representations of $\mathcal{G}$ to a direct sum~\cite{levy} (partial wave analysis).  One way of writing this direct sum reduction (including spin) is
\begin{eqnarray}\label{galCGC}
&&\HS(m_A,W_A,s_A)\otimes\HS(m_B,W_B,s_B)\\
&&=\int_{W=W_A+W_B}^\infty dW\bigoplus_{j=j_{min}}^\infty \HS(M,W,j)\otimes \mathbb{C}^{d(j,s_A,s_B)}.\nonumber
\end{eqnarray}
There is no sum over mass in the Galilean case, but there is a sum over internal (or center-of-mass) energy $W$ and intrinsic (or total center-of-mass) angular momentum $j$, where $j_{min}=0$ if both particles are either fermions or bosons and $j_{min}=1/2$ otherwise.  Since Galilean group is not simply reducible, the same UIR space $\HS(M,W,j)$ appears a number of times, $d(j,s_A,s_B)$.  
It is the number of ways orbital angular momentum $l$  combines with total spin $s$ to form total angular momentum $j$. The total spin in turn comes from the coupling of $s_A$ and $s_B$.

For particles with spin, the TPS (\ref{tps-tp4})  generalizes  as 
\begin{equation}\label{tps-tp5}
\HS(m_A,W_A,s_A)\otimes\HS(m_B,W_B,s_B)=\HS_\vP\otimes \HS_{int}
\end{equation}
where the internal Hilbert space is
\begin{equation}
\HS_{int} = \HS_W \otimes \bigoplus_{j=j_{min}}^\infty \HS(M,W,j)\otimes \mathbb{C}^{d(j,s_A,s_B)}
\end{equation}
We call this the internal-external (IE) TPS and will point out some of its special properties in the next section.

\section{Symmetry-Invariant TPSs}

Generally, a CSCO implies a TPS, and a transformation, often unitary, of the CSCO will lead to another TPS.  These TPSs may or may not be equivalent depending on whether the transformation factors into local unitary operators that act on each space in the original TPS.  In the case of interparticle entanglement, this notion of ``local'' coincides with more commonly acknowledged physical idea of spatially separated single particles.

An important class of transformations is symmetry transformations.  These generally arise as unitary representations $U$ of a group $G$ in the Hilbert space $\HS$ of states, $U:\ \HS\otimes G\rightarrow\HS$.   Now, consider the factorization $\HS=\HS_1\otimes\cdots\HS_n$  associated with a particular TPS. If the unitary representation $U$ of $G$ in $\HS$ factors $U=U_1\otimes\cdots U_n$ such that each $\HS_i$, $i=1,\cdots n$, is invariant under $U_i$ (and thus $U$), then we call that TPS invariant under the symmetry group $G$,  or more simply, symmetry invariant.  For example, the bipartite TPS $\HS=\HS_P\otimes\HS_Q$ would be symmetry-invariant under the representation $U$ of $G$, if, for all $g\in G$, $U(g)=U_Q(g)\otimes U_P(g)$.  {\em  Importantly, if a TPS is symmetry-invariant with respect to some group of transformations, then the notion of separability or entanglement is also invariant with respect to that symmetry.}

Returning to the example of the two different qubit TPSs in the first section, the group of rotations is generated by the total angular momentum operators which are proportional to $\hat{\sigma}^i_A\otimes\mathbb{I}_B + \mathbb{I}_A \otimes \hat{\sigma}^i_B$.  The unitary rotation operator is found by exponentiating this generator, and since $[\hat{\sigma}^i_A\otimes\mathbb{I}_B, \mathbb{I}_A \otimes \hat{\sigma}^i_B]=0$, this operator will be local with respect to the AB TPS.  In other words, for all $R\in\mathrm{SO}(3)$, we have $U(R)=U_A(R)\otimes U_B(R)$.  On the other hand, the PQ TPS is not symmetry invariant with respect to this representation of $\mathrm{SO}(3)$ (although a different representation could be constructed using PQ-local operators).  As a consequence, the amount of AB-type entanglement is independent of the orientation of the coordinate system, but PQ-type entanglement has some sort of directedness.

The idea of symmetry-invariant can be extended to particle systems by considering representations of the Galilean group.  If one uses the CSCO $\{ \hat{\bf P},\hat{S}_z\}$, the action of $U(g)$, $g=(b,{\bf{a}},{\bf{v}},R)\in\mathcal{G}$, on a vector $\phi_\chi(\vp)$ in $\HS(m,W,s)$  is~\cite{levy}
\begin{eqnarray}\label{uir1}
(U(g)\phi)_\chi(\vp) &=& e^{-i\frac{1}{2}m \mathbf{a}\cdot\mathbf{v} +i \mathbf{a}\cdot\vp' -i bE'} \nonumber\\
&&\times \sum_{\chi'}D^{s}({\mathcal{R}}(({\bf p}, E),\tilde{g}))_{\chi'\chi}\phi_{\chi'}(\vp')
\end{eqnarray}
where $\vp' = R\vp + m{\bf v}$, $E = 1/(2m)\vp^2 + W$, $E' = E + {\bf v}\cdot\vp + 1/(2 m){\bf v}^2$, $\tilde{g}=(0,0,{\bf{v}},R)$ 
and $\{{\mathcal{R}}(({\bf p}, E),{\tilde{g}})\}$ is an element of the ``little group" of $\mathcal{G}$ for a massive particle. 
The little group is the largest subgroup that leaves a standard momentum-energy pair  $({\bf{p}}_0,E_0)$ invariant.
For a massive particle, the little group of both the Galilean and Poincar\'e groups is isomorphic to the rotation group, and therefore, the 
$D^s({\mathcal{R}}(({\bf{p}},E),\tilde{g})$ is simply the unitary $2s+1$ dimensional representation of the rotation group. 
By definition, the little group, and therewith also the representation (\ref{uir1}), 
depend on the choice of $({\bf{p}}_0,E_0)$, which is arbitrary aside from the constraint $E-\frac{1}{2m}{\bf{p}}^2=W$. 
All of these different representations of $\mathcal{G}$ are equivalent, and therefore we may use any momentum-energy pair $({\bf{p}},E)$ to construct the general expression for the 
representation. The choice $({\bf{0}},W)$ is particularly simple in that $\{{\mathcal{R}}(({\bf 0}, W),{\tilde{g}})\}=R$, i.e., 
the little group of $\mathcal{G}$ can be chosen to be $SU(2)$, independently of the momentum and energy of the particle. 
With this choice, (\ref{uir1}) reduces to the simpler form
\begin{equation}
(U(g)\phi)_\chi(\vp) = e^{-i\frac{1}{2}m \mathbf{a}\cdot\mathbf{v} +i \mathbf{a}\cdot\vp' -i bE'} \sum_{\chi'}D^{s}(R)_{\chi'\chi}\phi_{\chi'}(\vp').
\label{uir1a}
\end{equation} 

From this we can see that under the standard representation of rotations for a single particle~\cite{levy}, the single-particle TPSs $\HS_\mathbf{p} \otimes \HS_{s}$ is rotation invariant, and with a little more work we can show that $\HS_E\otimes \HS_{L} \otimes \HS_s$ and $\HS_E\otimes \HS_{J}$ are also.  Therefore, a rotation (active or passive) will not change the associated forms of entanglement, even though the state itself may undergo a non-trivial transformation.

Further, we can see that the TPS $\HS_\mathbf{p} \otimes \HS_{s}$ is actually symmetry invariant under the full Galilean group because the unitary operators $U(g)$ factor into separate unitary operators $U(g)=U(g)_\vp\otimes U(g)_s$ acting on each Hilbert space.
Thus, intraparticle entanglement between the spin and the momentum of a free, non-relativistic particle is invariant across inertial reference frames.  We note that this is very different from the relativistic case, where the UIR does not factor and so momentum-spin entanglement is not invariant under coordinate transformations even for free particles~\cite{goodrel}. 

By considering representations on two particles, we can make similar identification of Galilean-invariant TPSs.  One that emerges is  are the interparticle TPS $\HS_A\otimes \HS_B$.  This can be seen because the product of two UIRs of the Galilean group will factor on the corresponding UIR spaces.  Combining this with the one-particle results, we find that the total momentum-total spin TPS $\HS_{\mathbf{p}_{tot}}\otimes\HS_{s_{tot}}$ is also symmetry invariant.

The IE TPS $\HS_\vP\otimes \HS_{int}$ is also invariant.  Proof of the this relies specifically on properties of the Clebsch-Gordan coefficients for the Galilean group (\ref{galCGC})~\cite{harshwick1}.  This result is satisfying because it agrees with intuition about bound states.  As discussed in \cite{ajp}, one expects that for the bound states of hydrogen, the internal DOF are unentangled with the external DOF, and that this result should not depend on the reference frame in any way.

\section{Dynamically Invariant TPSs and Conclusions}

One kind of symmetry transformation that  is of particular note is dynamical symmetry.  The Hamiltonian of a quantum system is the generator of time translations, the group of unitary operators that furnish a representation of the real line.  Therefore, it is relevant to ask the question which TPSs are invariant under time evolution. We call  such TPSs dynamical-invariant. For free particles, we have already identified dynamically invariant TPS because the free Hamiltonian is the generator of a subgroup of the Galilean group and all TPSs which are Galilean invariant are also dynamical invariant under free dynamics.

More interest is the case of interaction dynamics.  The general two-particle, the interaction-incorporating Hamiltonian $H$ that govern the dynamics cannot be expressed as a direct sum of  two commuting operators operators, say $H=H_A\otimes{\mathbb{I}}_B+{\mathbb{I}}_A\otimes H_B$ with respect to a given TPS. In particular, one would expect that for almost any Hamiltonian the interparticle TPS $\HS_A\otimes \HS_B$ would not be dynamical-invariant.  Nearly always, to interact is to entangle, as Sch\"odinger said years ago:
\begin{quotation}
When two systems...enter into temporary interaction..., and when they separate again, then they can no longer be described in the same way as before, viz.\ by endowing each of them with a representative of its own.~\cite{schro}
\end{quotation}
If the time evolution could be factored with respect to the interparticle TPS, then we could equivalently just redefine the internal energy operator $\hat{W}$.  Certain particle states will not be entangled  by certain interactions (see \cite{harshman_sym} for some examples), but generally interactions will generate entanglement with respect to the interparticle TPS.

While scattering is nearly always entangling with respect to the interparticle TPS, there is a TPS that is dynamically invariant for all Galilean invariant interactions, the IE TPS.  In \cite{harshwick1}, we prove this by showing that, as a consequence of  Schur's Lemma,  the S-matrix  must factor into unitaries on the IE TPS.  The applicability of Shur's lemma in turn relies on the fact that the S-matrix must commute with all generators of the Galilean Lie algebra, a necessary condition for non-relativistic interactions.  As an alternative justification, we note that for any interaction that depends only on relative positions and momenta, and not the total position or momenta, and that is rotationally invariant, the  total Hamiltonian can be decomposed into the direct sum
\begin{equation}
\hat{H} = \hat{H}_{ext}\otimes \mathbb{I} + \mathbb{I}\otimes \hat{H}_{int}\label{ham}.
\end{equation}
This condition is clearly met for all electromagnetic interactions, and it is more general than the restriction to central potential considered in 
\cite{harshman_sym}.  When (\ref{ham}) is exponentiated, the resulting one parameter unitary group will factor into local unitary operators with respect to the 
IE TPS. That is, the internal-external entanglement is dynamical-invariant.

In \cite{harshwick1}, we present an example of how this fact can be useful for understanding several recent results in \cite{spec1}.  We are currently investigating further applications and extensions of the above results to specific interactions, multiparticle systems, and relativistic systems.

Acknowledgments:  N.L.H. would like to thank the organizers of the 38th Symposium on Mathetmatical Physics in Toru\'n, Poland and is grateful for the support of the U.S.-Italian Fulbright Commission and the Research Corporation.

\end{document}